\documentclass[10pt,A4paper,conference]{IEEEtran}

\usepackage[pctex32]{graphics}
\usepackage{epsfig}
\usepackage{caption2}

\newcommand{\be}{\begin{equation}}
\newcommand{\ee}{\end{equation}}

\newcommand{\bea}{\begin{eqnarray}}
\newcommand{\eea}{\end{eqnarray}}

\newcommand{\bi}{\begin{itemize}}
\newcommand{\ei}{\end{itemize}}

\newcommand{\ben}{\begin{enumerate}}
\newcommand{\een}{\end{enumerate}}

\newcommand{\bef}{\begin{figure}[tbp]}
\newcommand{\enf}{\end{figure}}

\newcommand{\bt}{\begin{tabular}{lcllcl}}
\newcommand{\et}{\end{tabular}}

\newcommand{\bd}{\begin{description}}
\newcommand{\ed}{\end{description}}

\newtheorem{theorem}{Theorem}

\newcommand{\eref}[1]{(\ref{#1})}       % Equation reference.

\newcommand{\dfn}{\stackrel{\triangle}{=}}  % Equal by definition.
\newcommand{\comb}[2]{\left (
 \raisebox{-4pt}{$\stackrel{\mbox{\large $#1$}}{#2}$} \right ) }

\newcommand{\pvec}   {\mbox{\boldmath $\theta$}}

\newcommand{\sigvec} {\mbox{\boldmath $\sigma$}}

\newcommand{\etavec} {\mbox{\boldmath $\eta$}}
\newcommand{\xivec}  {\mbox{\boldmath $\xi$}}

\newcommand{\Beta}{{\cal B}}

\begin{document}

\title{Bounds on the Entropy of Patterns of I.I.D.\ Sequences}
%\title{Bounds on the Entropy of Patterns}

\author{\authorblockN{Gil I.\ Shamir\footnotemark{$^1$}}}
%\authorblockA{Electrical and Computer Engineering\\
%University of Utah \\
%Salt Lake City, UT 84112, U.S.A. \\
%e-mail: {\tt gshamir@ece.utah.edu}}}

\maketitle

%\footnotetext[1]{Supported in part by NSF Grant CCF-0347969.}
\footnotetext[1]{The author is with Department of Electrical and Computer
Engineering, University of Utah, Salt Lake City, UT 84112, U.S.A.,
e-mail:  gshamir@ece.utah.edu.
The work was partially supported by NSF Grant
CCF-0347969.}

\begin{abstract}
Bounds on the entropy of \emph{patterns\/} of sequences generated
by independently identically distributed (i.i.d.) sources are derived.
A pattern is a sequence of indices that contains all consecutive
integer indices in increasing order of first occurrence. If the
alphabet of a source that generated a sequence is unknown, the
inevitable cost of coding the unknown alphabet symbols can be
exploited to create the pattern of the sequence. This pattern can
in turn be compressed by itself.
The bounds derived here
are functions of the i.i.d.\ source entropy, alphabet size, and
letter probabilities.
It is shown that for large alphabets, the pattern entropy must decrease
from the i.i.d.\ one.  The decrease is in many cases more significant than
the universal coding redundancy bounds derived in prior works.
The pattern entropy is confined between two bounds that
depend on the arrangement of the letter probabilities in the
probability space.
For very large alphabets whose size may be greater than the coded pattern
length, all low probability letters are packed into one symbol.
The pattern
entropy is upper and lower bounded in terms of the i.i.d.\ entropy of the
new packed alphabet.
Correction terms, which are usually negligible,
are provided for both upper and lower bounds.

%{\bf Index Terms}: universal coding, patterns, index sequences,
%average redundancy, individual redundancy, minimax
%redundancy, maximin redundancy, redundancy for most sources,
%i.i.d.\ sources, MDL, redundancy-capacity theorem, sequential
%codes.
\end{abstract}

\section{Introduction}
\label{sec:introduction}

Several recent works (see, e.g., \cite{aberg97},
\cite{orlitsky04}-\cite{orlitsky04o},
\cite{shamir03}, \cite{shamir04c}, \cite{shamir04d})
have considered universal
compression for
\emph{patterns\/} of independently identically distributed
(i.i.d.)\ sequences. The pattern of a sequence is a sequence of
pointers that point to the actual alphabet letters, where the
alphabet letters are assigned \emph{indices\/} in order of first
occurrence. For example, the pattern of the sequence ``lossless''
is ``12331433''. A pattern sequence thus contains all positive
integers from $1$ up to a maximum value in increasing order of
first occurrence, and is also independent of the alphabet of the
actual data. Universal compression of patterns is interesting in
applications that attempt to compress sequences generated by an
initially unknown alphabet, such as a document in an unknown
language.  Utilization of the necessary coding of the unknown
symbols can take place by ordering the symbols in their order of
occurrence in the sequence, and then separately compressing the
alphabet independent pattern of the sequence.

To the best of our knowledge, universal compression of
patterns was first considered in
\cite{aberg97}, where it was proposed to compress sequences from a large
known alphabet in which not all symbols are expected to occur by separating the
representation of the occurring
alphabet symbols from the pattern and compressing each separately.  The paper
considered compression of individual sequences.
Later, patterns were rediscovered (and named) in a series of papers
\cite{orlitsky04}, \cite{orlitsky04o} (and references therein) that considered
universal compression for unknown alphabets and thoroughly studied
the redundancy in universal coding of
individual pattern sequences.  These
papers demonstrated that the individual sequence redundancy of patterns
must decrease in universal
compression compared to the redundancy obtained for
simple universal compression of i.i.d.\ sequences.
%\cite{kieffer78}, \cite{orlitsky04o}, \cite{shamir04}.
Furthermore, unlike the i.i.d.\ case,
%\cite{kieffer78},
it was shown that
the redundancy in universal pattern compression vanishes
even if the alphabet is infinite.
(This is, of course, related to the fact that we loose some information
by coding the pattern instead of the actual sequence.)
The universal average case was then studied in \cite{shamir03}, \cite{shamir04c},
\cite{shamir04d}, where redundancy bounds for average case universal compression
of patterns were derived.
%Additional work on patterns included work on
%distribution modeling \cite{orlitsky04ciss}.

The universal description length of patterns, however, consists of the pattern
entropy and the redundancy of universally coding the pattern.  While most of the
emphasis in prior work was on the latter, it is clear that a pattern is a data
processing over the actual sequence, and thus its entropy (the first term) must decrease.
Furthermore, in \cite{shamir04d} (see also \cite{shamir03}), we derived sequential codes
for compressing patterns and bounded their description length.  It was shown that for
sufficiently large alphabets
this description length was significantly smaller than the i.i.d.\ source
entropy.  This points out to the fact that not only is there an entropy decrease
in patterns, but for large alphabets, this decrease is much more significant than
the increase in description length due to the universal redundancy.  Hence, to have
better understanding even of universal compression of patterns, it is essential also
to study the behavior of the pattern entropy.  Pattern entropy is also important in
learning applications.   Consider all the new faces that a newborn sees.  The newborn
can identify these faces with the first time each was seen.  There is no difference
if it sees nurse $A$ or nurse $B$ (and never sees the other),
as long as it is a nurse.  The entropy of patterns
can thus model the uncertainty of such learning processes.
The exponent of the entropy gives an approximate count of the typical patterns
one is likely to observe for the given source distribution.
If the uncertainty goes to $0$, we are likely to observe only one pattern.

We first considered pattern entropy in \cite{shamir03a}, where we
bounded the range of values within which
the entropy of a pattern can be, depending on the specific
distribution, as a function of the i.i.d.\ entropy.
We showed that
for larger alphabets, the pattern entropy \emph{must\/} decrease
with respect to
(w.r.t.)\ the i.i.d.\ one.  However, the results were limited to
distributions that contain only letters with sufficiently large letter
probability.  An upper bound that extends the results for unbounded distributions
was derived in \cite{shamir04a}.  Subsequently to our initial paper \cite{shamir03a},
pattern entropy was independently
studied with different approaches and from a different view
of the problem in \cite{gemelos04} and
\cite{orlitsky04itw}, where the focus has been
on limiting results for the entropy \emph{rate\/} of patterns.

In this paper, we continue and generalize the results in \cite{shamir03a} and
\cite{shamir04a}.  We derive general upper and lower
bounds for the entropy of patterns generated by large and very large alphabets.
The bounds are presented as functions of a related i.i.d.\ entropy,
the alphabet size, and the alphabet letter probabilities.  The related i.i.d.\
entropy is that of the i.i.d.\ source if no letters with very low probabilities
exist.  Otherwise, all the probabilities smaller than a threshold are packed into
one symbol, and the i.i.d.\ entropy is that of the new alphabet.
Since the detailed proofs of most of the bounds require lengthy rigorous analysis,
we only include road maps of the proofs in this paper.  The complete proofs
are presented in \cite{shamir05}.

The technique used to derive the bounds in this paper relies on partitioning the
probability space into a \emph{grid\/} of points.  Between each two points, we obtain
a \emph{bin\/}.  For a typical i.i.d.\ sequence of the source, each
permutation of the sequence letters that only permutes among letters in the same
bins, has almost the same probability as the typical sequence, and results in the
same pattern.  Such permutations exchange all occurrences of one letter by
all occurrences of another.
The probability of the pattern increases from that of the
i.i.d.\ sequence by the number of such permutations.  This, in turn, yields a decrease
in the pattern entropy.  This idea is used directly to derive some of the
bounds, and is extended to include low probabilities to derive the more general
bounds.  To derive a general upper bound, we
propose a low-complexity sequential (non-universal) code for
compressing patterns, which achieves the bound.  The algorithm is,
again, based on the idea of bins.  The use of bins is not easy because the grids
that determine the bins need to be wisely designed to efficiently utilize the
probability space.  In particular, the grid points are taken in increasing spacing.
The reason is that for large probabilities, the decrease in probability assigned
to a typical sequence is slower as we shift away from the true letter probability.

The outline of the paper is as follows.
In Section~\ref{sec:note_def}, we define the notation.
Section~\ref{sec:background} reviews initial simple, easy to derive, bounds on the entropy,
and motivates the remainder of the paper.
Then, in Section~\ref{sec:large}, we derive the upper and lower bounds
for pattern entropy of i.i.d.\ sources with sufficiently large probabilities, and
show the range of values that the pattern entropy can take in this case, depending
on the actual source distribution.
Finally, Section~\ref{sec:very} contains the derivations of more general
upper and lower bounds, that do not require a condition on the letter probabilities.

\section{Notation and Definitions}
\label{sec:note_def}

Let $x^n \dfn \left ( x_1, x_2, \ldots, x_n \right )$ be a sequence
of $n$ symbols over an alphabet $\Sigma$
of size $k$.
The parameter $\pvec \dfn \left ( \theta_1, \theta_2, \ldots, \theta_k \right )$
contains the probabilities of the alphabet letters.
Since the order of these probabilities does not affect the
pattern, we assume, without loss
of generality, that $\theta_1 \leq \theta_2 \leq \cdots \leq \theta_k$, and that
$\Sigma  =  \left \{ i, 1 \leq i \leq k \right \}$.
In general, boldface letters will denote vectors, whose components will
be denoted by their indices.
Capital letters will denote random variables.

The \emph{pattern\/} of $x^n$ will be denoted by
$\psi^n \dfn \Psi \left ( x^n \right )$.  Different sequences have the same pattern.
For example, for
the sequences $x^n =$``lossless'', $x^n =$``sellsoll'',
$x^n =$``12331433'', and $x^n =$``76887288'', the pattern is
$\Psi \left ( x^n \right ) =$``12331433''.  Therefore, for given
$\Sigma$ and $\pvec$, the probability of a pattern $\psi^n$ \emph{induced\/}
by an i.i.d.\ underlying probability is given by
\be
\label{eq:pattern_probability}
 P_{\theta} \left ( \psi^n \right ) =
  \sum_{y^n: \Psi (y^n) = \psi^n } P_{\theta} \left ( y^n \right ).
\ee
The probability of $\Psi \left ( x^n \right )$ can be expressed as
in \eref{eq:pattern_probability} by summing over all sequences that
have the same pattern with a fixed parameter vector.
However, we can also express it
by fixing the actual sequence and summing over all permutations of
occurring symbols of the parameter vector
\be
 \label{eq:pattern_prob1}
 P_{\theta} \left [ \Psi \left ( x^n \right ) \right ] =
 \sum_{\sigvec} P_{\theta(\sigma)} \left (x^n \right ),
\ee
where the summation is over all permutation
vectors $\sigvec$ that differ among each other
in the index of the probability parameter assigned to at least
one occurring letter, and
$\theta \left ( \sigma_i \right )$ denotes the $i$th component of the
permuted vector $\pvec$, permuted according to $\sigvec$.
For example,
if $\pvec = \left ( 0.7, 0.1, 0.2 \right )$ and
$\sigvec = \left (3, 1, 2 \right )$,
then $\pvec \left ( \sigvec \right ) = \left ( 0.2, 0.7, 0.1 \right )$
and $\theta \left ( \sigma_2 \right ) = \theta_1 = 0.7$.

The entropy rate of an i.i.d.\ source will be denoted by
$H_{\theta} \left (X \right )$.  The sequence entropy for an i.i.d.\
source is $H_{\theta} \left (X^n \right ) = n H_{\theta} \left (X \right )$.
The
\emph{pattern sequence entropy\/} of order $n$ of a source $\pvec$ is defined
as
\be
\label{eq:pattern_entropy}
 H_{\theta} \left ( \Psi^n \right ) \dfn
 -\sum_{\psi^n} P_{\theta} \left ( \psi^n \right )
 \log P_{\theta} \left ( \psi^n \right ).
\ee

As described in Section~\ref{sec:introduction}, we will grid the
probability space in order to derive the bounds.
Letters whose probabilities lie in
the same bin between two adjacent grid points
will be grouped together.  We will use two different grids, as defined below,
to derive the bounds.
For an arbitrarily small $\varepsilon > 0$, let $\etavec \dfn \left (\eta_0, \eta_1,
\eta_2, \ldots, \eta_b, \ldots, \eta_{B} \right )$ be a grid of $B+1$ points, where
$\eta_0 = 0$, $\eta_1 = 1/n^{1+\varepsilon}$, and let,
\be
 \label{eq:tau_grid_def}
 \tau'_b \dfn
 \sum_{j=1}^{b} \frac{2 (j - \frac{1}{2})}{n^{1+2\varepsilon}} =
 \frac{b^2}{n^{1+2\varepsilon}}.
\ee
Then,
\be
 \label{eq:eta_grid_def}
 \eta_b = \tau'_{b+n^{3\varepsilon/2} - 2};~~\forall b \geq 2,
\ee
i.e., $\eta_2 = 1/n^{1-\varepsilon}$, and so on.
Clearly, there are $B$ nonzero grid points, where $B$ is the rounded down
integer of $\sqrt{n}^{1+2\varepsilon} - n^{3\varepsilon/2} + 2$.

We will use $k_b$ to denote the number of letters
$\theta_i \in \left ( \eta_b, \eta_{b+1} \right ]$.  In particular, $k_0$
will denote the number of letters in $\Sigma$ with probability
not greater than $1/n^{1+\varepsilon}$, $k_1$ the number of
letters with probabilities in $\left ( 1/n^{1+\varepsilon},
1/n^{1-\varepsilon} \right ]$, and $k_{01}$ their sum.
Let $\varphi_b$
be the total probability of letters in bin $b$ of grid
$\etavec$. Of particular importance will be $\varphi_0$,
$\varphi_1$, defined w.r.t.\ bins $0$, $1$, respectively, and
$\varphi_{01} \dfn \varphi_0 + \varphi_1$.  We use $L$, and $L_b$
for the mean number of total letters, and letters from bin $b$,
respectively, that occur in $X^n$, i.e.,
%\be
% \label{eq:mean_bin}
$ L_b = \sum_{\theta_i \in \left ( \eta_b, \eta_{b+1} \right ]}
 \left [ 1 - \left ( 1 - \theta_i \right )^n \right ]$.
% = k_b - \sum_{\theta_i \in \left ( \eta_b, \eta_{b+1} \right ]}
% \left ( 1 - \theta_i \right )^n
%\ee
It is easy to see that
\be
 \label{eq:mean_bin_bound}
 k_b - \sum_{\theta_i \in \left ( \eta_b, \eta_{b+1} \right ]}
 e^{-n \theta_i} \leq
 L_b \leq
 k_b - \sum_{\theta_i \in \left ( \eta_b, \eta_{b+1} \right ]}
  e^{-n \left (\theta_i-\theta_i^2 \right )},
\ee
where in the upper bound summation
only $\theta_i \leq 3/5$ are included.  In particular, for bin $0$,
\be
 \label{eq:min_bin0_bound}
 n \varphi_0 - \comb{n}{2} \sum_{i=1}^{k_0}
 \theta_i^2 \leq L_0 \leq
 n \varphi_0 - \comb{n}{2} \sum_{i=1}^{k_0}\theta_i^2 +
 \comb{n}{3} \sum_{i=1}^{k_0}\theta_i^3.
\ee

The points $b \geq 1$ in
grid $\xivec \dfn \left ( \xi_0, \xi_1, \ldots, \xi_{\bar{B}} \right )$
are defined as in \eref{eq:tau_grid_def}, but where $-\varepsilon$ replaced
$2 \varepsilon$, and also $\xi_0 = 0$.
Here, we will use $\kappa_b$, $b\geq 1$, to denote the number of letters whose probabilities
are in the
three adjacent bins surrounding $b$, i.e.,
$\theta_i \in \left (\xi_{b-1}, \xi_{b+1} \right ]$, with the exception of $\kappa_1$
which will only count the letters with probabilities
in $\left (\xi_1, \xi_2 \right ]$.

Using the definitions above, we can now define two i.i.d.\ entropy expressions,
where some of the low probability symbols are packed into one symbol,
\bea
% \label{eq:zero_bin_packed_entropy}
% H_{\theta}^{(0)} \left ( X \right ) &\dfn&
% -\varphi_0 \log \varphi_0 - \sum_{i=k_0+1}^k \theta_i \log \theta_i, \\
 \label{eq:zeroone_bin_packed_entropy}
 H_{\theta}^{(01)} \left ( X \right ) &\dfn&
 -\varphi_{01} \log \varphi_{01} - \sum_{i=k_{01}+1}^k \theta_i \log \theta_i, \\
 \label{eq:zero_one_bin_packed_entropy}
 H_{\theta}^{(0,1)} \left ( X \right ) &\dfn&
 -\sum_{b=0}^1 \varphi_b \log \varphi_b -
% -\varphi_0 \log \varphi_0 - \varphi_1 \log \varphi_1 -
 \sum_{i=k_{01}+1}^k \theta_i \log \theta_i.
\eea

\section{Background and Simple Bounds}
\label{sec:background}

It is clear that the pattern entropy satisfies the following bounds:
\begin{theorem}
\label{theorem:simple_bounds}
If $k \leq n$,
\be
 \label{eq:entropy_bounds_simple1}
 nH_{\theta} \left (X \right ) - \log \left ( k! \right ) \leq
 H_{\theta} \left ( \Psi^n \right ) \leq
 nH_{\theta} \left (X \right ).
\ee
Otherwise,
\be
 \label{eq:entropy_bounds_simple2}
 nH_{\theta} \left (X \right ) - \log \frac{k!}{(k-n)!} \leq
 H_{\theta} \left ( \Psi^n \right ) \leq
 nH_{\theta} \left (X \right ).
\ee
\end{theorem}
The upper bound is trivial, and the
lower bounds are proved in \cite{shamir05}.  For $k = o(n)$, the simple bound
in \eref{eq:entropy_bounds_simple1} already points the fact that if the i.i.d.\
entropy rate of the source is not vanishing, the entropy
rate of patterns is equal to the i.i.d.\ one.  However, it is clear that
for many sources the bounds above are not tight.

In \cite{shamir04d}, we derived a universal
sequential algorithm for coding patterns.
The bound on its description length provides
a bound on the pattern entropy.  In particular, if $\hat{k} > e^{19/18}\cdot n^{1/3}$,
where $\hat{k}$ is the number of alphabet letters that occur in $x^n$ with
probability at least $(1 - \varepsilon)$,
it was shown that the pattern entropy must decrease from the i.i.d.\
one, and
\be
 \label{eq:entropy_bound_dcc}
 H_{\theta} \left ( \Psi^n \right )
 \leq
 n H_{\theta} \left ( X \right ) -
 \left ( 1 - \varepsilon \right )
 \frac{3}{2} \hat{k} \log \frac{\hat{k}}{e^{19/18} n^{1/3}}.
\ee
For sources with very high entropy, for example,
$\theta_i = n^{-\alpha}, \forall i$, for some constant $\alpha
\geq 1$, the bound increases with $n$ and becomes loose.
The derivation in \cite{shamir04d} can thus be used to replace the
first term in the bound by $n \log n$.
However, this still yields
a very loose bound on the entropy.

\section{Bounds for Small and Large Alphabets}
\label{sec:large}

We now consider sources in which $\theta_1 > 1/n^{1-\varepsilon}$, i.e.,
$k_{01} = 0$.  We
present an upper bound and a lower bound for this case,
and discuss the range of values the entropy can
take, where for sufficiently large $k$, it
\emph{must\/} decrease from the i.i.d.\ one.

\subsection{An Upper Bound}

The following theorem upper bounds the pattern entropy.
\begin{theorem}
\label{theorem:ub2}
Let $\theta_i > 1/n^{1-\varepsilon}$, $\forall i, 1\leq i \leq k$.  Then,
\be
 \label{eq:ub2}
 H_{\theta} \left ( \Psi^n \right ) \leq
 n H_{\theta} \left ( X \right ) -
 \left ( 1 - \varepsilon \right )
 \sum_{b=2}^{B} \log \left ( k_b ! \right ).
\ee
\end{theorem}

To prove Theorem~\ref{theorem:ub2}, we lower bound the probability of patterns
generated only from typical sequences $x^n$ by the
sum of probabilities of all typical sequences that have
this pattern.
Using \eref{eq:pattern_prob1} and this idea,
$P_{\theta} \left [ \Psi \left ( x^n \right ) \right ]$
is lower bounded by the partial sum of permutations $\sigvec$ of $\pvec$, that
contains only permutations for which for every $i$ and every $b$,
$\theta_i \in \left ( \eta_b, \eta_{b+1} \right ]
\Rightarrow \theta \left ( \sigma_i \right ) \in \left ( \eta_b, \eta_{b+1} \right ]$.
For all such permutations and a typical sequence $x^n$, the probability
assigned to $x^n$ decreases at most negligibly w.r.t.\ the actual probability
of $x^n$.
Hence, for a typical $x^n$,
\be
 \label{eq:ub1_p1}
 \log P_{\theta} \left [ \Psi \left ( x^n  \right ) \right ] \geq
 \log P_{\theta} \left ( x^n \right ) +
 \log M_{\theta} - o(k),
\ee
where $M_{\theta}$ is the number of such permutations $\sigvec$.
Computing $M_{\theta}$ and accounting for the probability of non-typical sequences
yields the bound of \eref{eq:ub2}.

\subsection{A Lower Bound}

The next theorem shows a bound of similar nature to
the bound of Theorem~\ref{theorem:ub2}.
\begin{theorem}
\label{theorem:lb1}
Let $\theta_i > 1/n^{1-\varepsilon}$, $\forall i, 1\leq i \leq k$.  Then,
\be
 \label{eq:lb1}
 H_{\theta} \left ( \Psi^n \right ) \geq
 n H_{\theta} \left ( X \right ) -
 \sum_{b=1}^{\bar{B}} \log \left ( \kappa_b ! \right ) - o(1).
\ee
\end{theorem}

To prove Theorem~\ref{theorem:lb1} we first define a typical pattern $\psi^n$ as one that
is the pattern of at least one typical $x^n$.   The number of typical
sequences $x^n$ that have a given typical pattern is then upper bounded by the product
of factorials that leads to the second term of the bound.  It is then shown
that the contribution of non-typical sequences to the probability
of any typical pattern decays exponentially in $n^{\alpha \varepsilon}$, where
$\alpha$ is some constant.
It is necessary to show that even if a typical pattern is the pattern of very few
typical sequences, the many non-typical sequences of this pattern still
contribute negligibly to its probability.  To show that,
each set of non-typical sequences that have pattern $\psi^n$ is shown to
result from a permutation
of a typical sequence, where the probability of such a non-typical permutation multiplied
by a bound on the number of such permutations is still negligible w.r.t.\ the
probability of the original typical sequence.  Finally, a straightforward
set of equations that breaks the pattern entropy computation into typical
and non-typical sequences, yields the bound of \eref{eq:lb1}.

\subsection{Entropy Range}

We now consider the overall range of values the pattern entropy can take,
regardless of how the letter probabilities are lined
up in the probability space.
It is clear that the lower
bound in \eref{eq:entropy_bounds_simple1} is tight for a uniform distribution
for $\theta_i > 1/n^{1-\varepsilon}$.
The upper bound, however, is restricted by the minimum number
of permutations that yield a typical sequence after permuting another typical
sequence.  For the simple bound in \eref{eq:entropy_bounds_simple1}, only
the identity permutation is counted.  However, if the number of alphabet
symbols is sufficiently large, there \emph{must\/} be more than one such permutation,
because more than one letter probability must fall within a single bin
of $\etavec$.  Letters with probabilities in the same bin in a typical
$x^n$ can be permuted among themselves to another sequence $y^n$ that
is typical, gives the same pattern,
and has almost equal probability to $x^n$.
Not to violate the condition $\sum \theta_i = 1$, most of the letter probabilities
must be distributed in
essentially $O \left ( n^{(1+\varepsilon)/3} \right )$ lower bins of $\etavec$.
For sufficiently large alphabets, using the smallest possible number of
such permutations, yields
\begin{theorem}
\label{theorem:range}
Let $\theta_i > 1/n^{1-\varepsilon}$, $\forall i, 1\leq i \leq k$, and
let $k \geq n^{(1+\varepsilon)/3}$.  Then,
\bea
 \nonumber
  \lefteqn{n H_{\theta} \left ( X \right ) - \log \left (k! \right ) \leq
   H_{\theta} \left ( \Psi^n \right )
  }\\
  \label{eq:range}
  &\leq&
  n H_{\theta} \left ( X \right ) -
  \left ( 1 - \varepsilon \right )
  \frac{3}{2} k \log \frac{k}{e^{2/3} n^{1/3}}.
\eea
\end{theorem}

Theorem~\ref{theorem:range}
gives a range within which the pattern
entropy must be, depending on the actual letter probabilities.
Figure~\ref{fig:entropy ratio} shows
the region of decrease in the pattern entropy w.r.t.\
the i.i.d.\ one.  For large alphabets, the entropy must decrease essentially
by at least $1.5 \log \left (k / n^{1/3} \right )$ bits per
alphabet symbol.
\bef
\centerline{\includegraphics
%[height=7cm,width=8.5cm]
[bbllx=50pt,bblly=185pt,bburx=560pt,
bbury=598pt,height=5cm, width=7cm, clip=]{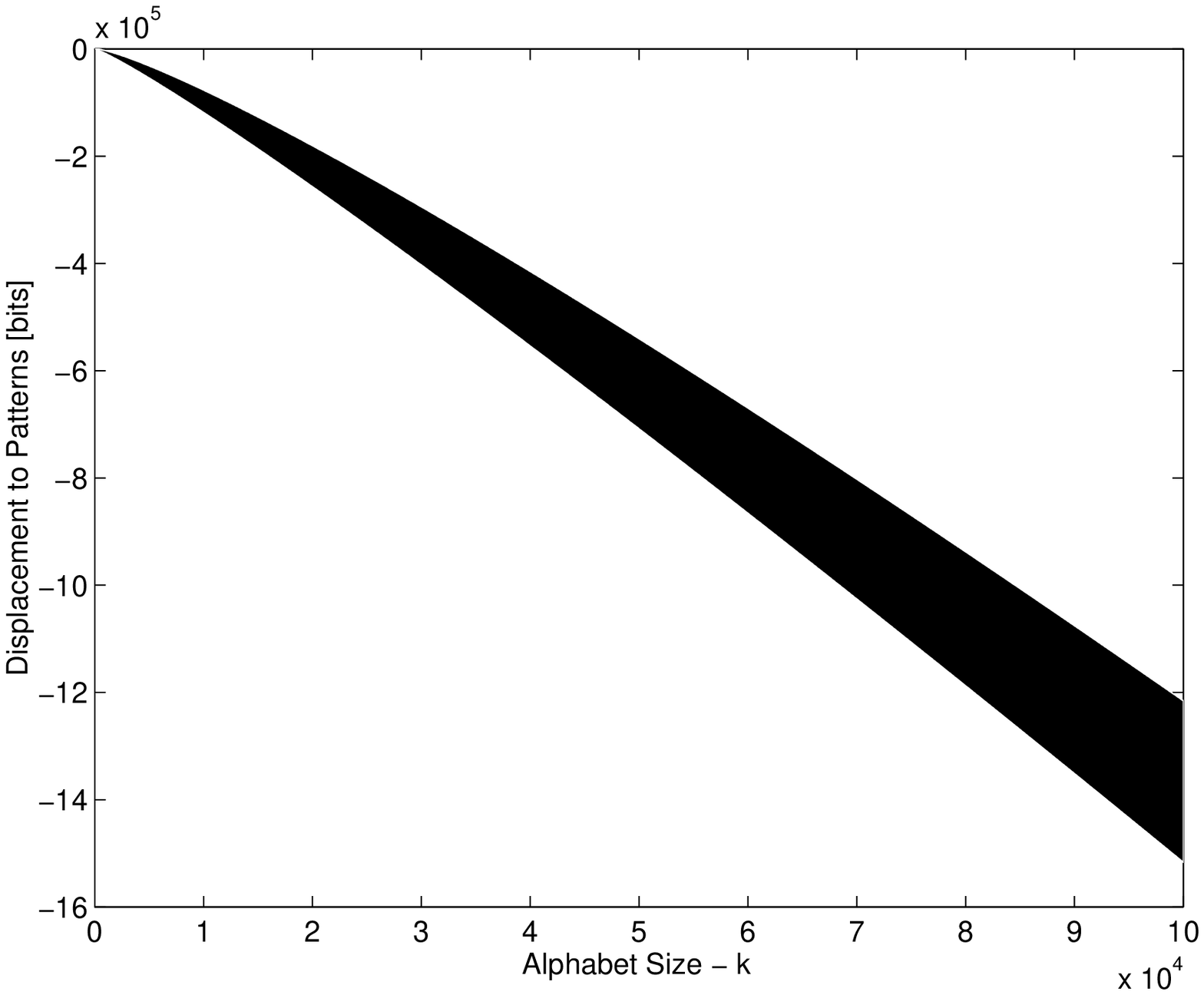}}
\caption{Region of decrease from i.i.d.\ to pattern entropy as function
of $k$ for $n=10^6$ bits with $\varepsilon = 0.1$.}
\label{fig:entropy ratio}
\enf

\section{Bounds for Very Large Alphabets}
\label{sec:very}

We now consider a more general case, where there is no
lower bound on the letter probabilities.

\subsection{An Upper Bound}

A general upper bound on $H_{\theta} \left ( \Psi^n \right )$ is
derived through a sequential
probability assignment code.  A new symbol is assigned
a joint probability of its index and its bin in the grid $\etavec$.  We thus
code the joint sequence $\left (\psi^n, \beta^n \right )$,
where $\beta^n$ is the sequence of bin indices corresponding to $x^n$. The average
description length of this code upper bounds the joint entropy
$H_{\theta} \left ( \Psi^n, \Beta^n \right )$, which in turn upper bounds
$H_{\theta} \left (\Psi^n \right )$.

The probability that is assigned to the joint pattern and bin sequence is given
by
%\be
% \label{eq:seq_prob_total}
$ Q \left [ \left (\psi^n, \beta^n \right ) \right ] \dfn
 \prod_{j=1}^n Q \left [ \psi_j, \beta_j ~|~ \left ( \psi^{j-1},
 \beta^{j-1} \right ) \right ]$.
%\ee
If $\psi_j$ is an index that already occurred in the pattern $\psi^{j-1}$,
then
\be
\label{eq:seq_prob_old}
  Q \left [ \psi_j, \beta_j ~|~ \left ( \psi^{j-1}, \beta^{j-1} \right ) \right ] =
  \rho_{\beta_j},
\ee
where $\rho_b \dfn \varphi_b / k_b$ for $b \geq 2$, and $\rho_0$ and $\rho_1$ are
values assigned to letters in the first two bins, that will be optimized later.
%to minimize the average description length over all patterns.
Once an index
occurred, it only occurs jointly with
the same bin number that occurred with its first occurrence.
If $\psi_j$ is a new index, and its bin
is $\beta_j$, the pair is assigned probability
\be
\label{eq:seq_prob_new}
  Q \left [ \psi_j, \beta_j ~|~ \left ( \psi^{j-1}, \beta^{j-1} \right ) \right ] =
  \varphi_{\beta_j} - c\left [\left (\psi^{j-1}, \beta^{j-1} \right ), \beta_j \right ]
   \cdot \rho_{\beta_j},
\ee
where $c\left [\left (\psi^{j-1},\beta^{j-1} \right ), \beta_j \right ]$
is the number of distinct indices that jointly occurred with
bin index $\beta_j$ in $\left (\psi^{j-1}, \beta^{j-1} \right )$ (e.g., if
$\psi^{j-1} = 1232345$ and $\beta^{j-1} = 1222242$ then
$c \left [ \left (\psi^{7}, \beta^{7} \right ), \beta_j \right ]$ is
$3$ for $\beta_j = 2$, $1$ for $\beta_j = 1$ and $\beta_j = 4$, and
is $0$, otherwise).

This probability assignment
groups the probability of
all the symbols in the same bin into one symbol.  Then,
each occurrence of a new symbol in bin $b$, it codes a new index with the remaining
group probability, extracting one count of the mean bin probability from the
remaining probability in the bin.
Each re-occurrence of an index assigns the index and its attached bin
the mean bin probability of the respective bin.
%This reflects assignment of the mean bin probability
%to each re-occurrence of a symbol whose probability is within any given bin.
For bins $b=0,1$, the mean is replaced by $\rho_0$ and $\rho_1$,
respectively.

Upper bounding the average description length of this code,
optimizing $\rho_0$ and $\rho_1$ to minimize the bound, yields
the following upper bound on the pattern entropy.
\begin{theorem}
\label{theorem:ub3}
The pattern entropy is upper bounded by
\bea
 \nonumber
 H_{\theta} \left ( \Psi^n \right ) &\leq&
 n H^{(0,1)}_{\theta} \left ( X \right ) -
 \sum_{b=2}^B \left ( 1 - \varepsilon \right )
 \log \left ( k_b ! \right ) \\ &+&
 \nonumber
 \left ( n \varphi_1 - L_1 \right ) \log
 \left [ \min \left \{ k_1, n \right \} \right ] +
 n \varphi_1 h_2 \left ( \frac{L_1}{n\varphi_1} \right ) \\ &+&
 \label{eq:ub3}
 \left ( \frac{n^2}{2} \sum_{i=1}^{k_0} \theta_i^2 \right )
 \log  \left \{ \frac{2 e \cdot \varphi_0 \cdot \min \left \{k_0, n \right \}}
 {n \sum_{i=1}^{k_0} \theta_i^2} \right \},
\eea
where $h_2 \left ( \alpha \right ) \dfn
-\alpha \log \alpha - (1-\alpha) \log (1 - \alpha)$.
\end{theorem}
The bound consists of:  the
packed i.i.d.\ entropy with bins $0$ and $1$ as one symbol each
(the first term),
the pattern gain in first occurrences of any
letter within the
remaining bins (the second term), the loss in packing bin $b=1$ (the next
two terms), and the loss in packing bin $b=0$ (the last term).  The
greatest contribution of the third and the fourth term can be shown to be
$(1 -\varepsilon )n \varphi_1 \log n$, and that
of the last
term $0.5 \varphi_0 n^{1-\varepsilon} \log \left ( 2 e n^{1+\varepsilon} \right )$,
which is clearly negligible if
$H^{(0,1)}_{\theta} \left ( X \right )$ is non-vanishing.
%We also note that the lower bound on $L_1$ from \eref{eq:mean_bin_bound} can
%replace $L_1$ in the bound as long as $k_1 \geq (1 +\varepsilon)n^{\varepsilon}$.
%Otherwise, the lower bound can still be used for the third term, and the proper
%bound for the fourth.

\subsection{A Lower Bound}

To lower bound $H_{\theta} \left ( \Psi^n \right )$, the contributions
of large and small probabilities are separated.  The former, of probabilities
greater than $1/n^{1-\varepsilon}$, is bounded using
derivation as in
Theorem~\ref{theorem:lb1}.  The latter is bounded by a straightforward derivation.
To separate the two, we define a random sequence $Z^n$, such that
$Z_j = 0$ if $\theta_{x_j} \leq 1/n^{1-\varepsilon}$ and $1$ otherwise.  Using $Z^n$,
$H_{\theta} \left ( \Psi^n \right )$ can be expressed as
\be
 \label{eq:z_entropy}
 H_{\theta} \left ( \Psi^n \right ) =
 H_{\theta} \left ( \Psi^n ~|~ Z^n \right ) +
 H_{\theta} \left ( Z^n \right ) -
 H_{\theta} \left ( Z^n ~|~ \Psi^n \right ).
\ee
The first term of \eref{eq:z_entropy} can now be bounded by splitting a particular
value $z^n$ of $Z^n$ into the elements for which $z_j = 1$ and those for which
$z_j = 0$, and bounding $H_{\theta} \left ( \Psi^n ~|~ z^n \right )$ separately
for each of these sets.  We use the relation
$H_{\theta} \left ( \Psi^n ~|~ z^n \right ) =
\sum_{j=1}^n H_{\theta} \left ( \Psi_j ~|~ \Psi^{j-1}, z^n \right )
\geq \sum_{j=1}^n H_{\theta} \left ( \Psi_j ~|~ X^{j-1}, z^n \right )$.
%\bea
% \nonumber
% H_{\theta} \left ( \Psi^n ~|~ z^n \right ) &=&
% \sum_{j=1}^n H_{\theta} \left ( \Psi_j ~|~ \Psi^{j-1}, z^n \right ) \\
% &\geq&
% \label{eq:pattern_processing}
% \sum_{j=1}^n H_{\theta} \left ( \Psi_j ~|~ X^{j-1}, z^n \right ).
%\eea
%The right hand side of \eref{eq:pattern_processing} is much easier to compute.
%The second term of \eref{eq:z_entropy} is trivial.  However,
The third term of \eref{eq:z_entropy}
complicates
the analysis if there is no clear separation between small and
large probabilities, i.e., there exits $\varepsilon$ values for which
there are $k^-_2 > 0$ letters with probabilities
in $\left ( 1/ (2n^{1-\varepsilon}), 1/ n^{1-\varepsilon} \right ]$ and
$k^+_2 > 0$ letters with
probabilities
in $\left ( 1/ n^{1-\varepsilon}, 3/ (2n^{1-\varepsilon}) \right ]$.  A permutation
between letters in the first bin and letters in the second may still result in a
typical sequence.  Hence, the separation must yield a correction term.
%for such sources, for any choice
%of arbitrarily small $\varepsilon$.
%(Note that in some sources, but not all, it may
%be possible to choose $\varepsilon$ that bypasses this problem.)
Applying all the above considerations yields the following lower bound:
\begin{theorem}
\label{theorem:lb2}
The pattern entropy is lower bounded by
\bea
 \nonumber
 H_{\theta} \left ( \Psi^n \right ) &\geq&
 n H^{(01)}_{\theta} \left ( X \right ) -
 \sum_{b=1}^{\bar{B}} \log \left ( \kappa_b ! \right ) \\
 \nonumber
 &+&
 \sum_{i=1}^{k_{01} - 1} \left [ n \theta_i - 1 + e^{-n \left ( \theta_i +
 \frac{\theta_i^2}{\varphi_{01}} \right )} \right ]
 \log \frac{\varphi_{01}}{\theta_i} \\
 \nonumber
 &+&
 \left ( n \theta_{k_{01}} - 1 \right ) \log \frac{\varphi_{01}}{\theta_{k_{01}}} \\
 \nonumber
 &+&
 (\log e) \sum_{i=1}^{L_{01}-1} \left ( L_{01} - i \right )
 \frac{\theta_i}{\varphi_{01}} \\
 &-&
 \log \comb{k^-_2 + k^+_2}{k^+_2} - o(1).
 \label{eq:lb2}
\eea
\end{theorem}
The first term in \eref{eq:lb2} is the i.i.d.\ entropy in which all letters
with probability not greater than $1/n^{1-\varepsilon}$ are packed into one symbol.
The second term is the decease in entropy due to first occurrences of large
probability letters.  The next three terms are due to the contribution of low
probability letters beyond that of the super-symbol that merges them.  The first
two of these represent the penalty in packing in repetition of these letters, where
the third one is the penalty in first occurrence of such a letter.  We note that
the first
two of these three terms can be separated into contributions of the first $k_0$ letters
and the following $k_1$ letters, to obtain expressions that resemble the
bound in \eref{eq:ub3}.  The sixth term of \eref{eq:lb2} is the correction term from
separating small and large probability letters.  Finally, the last term of
$o(1)$ absorbs all
the lower order terms of the bound.  As in the upper bound in \eref{eq:ub3},
all the terms beyond the first two and the first element of the third term,
can be shown to contribute at most $O \left (n \varphi_1 \log n \right )$ for
letters that result from bin $1$ of $\etavec$, and $o(n)$ for letters that
result from bin $0$ of $\etavec$.

There are several other forms that the bound in \eref{eq:lb2} can be brought to.
In particular, the second term, representing the decrease due to first occurrences
of large probability letters may not be tight if the distribution is close to uniform,
but symbols appear in very few separate adjacent bins formed by $\xivec$.  If this
is the case, it may be beneficial to derive a bound on the large probabilities
using similar methods to the bound derived on the low probabilities.  In such a
bound, the second term of \eref{eq:lb2} will be replaced by two terms that take
the form of the third and fifth terms of \eref{eq:lb2}, where the $n\theta_i$
leading element of the third term is omitted.

\section{Summary and Conclusions}

We studied the entropy of patterns of i.i.d.\ sequences.
We provided upper and lower bounds on this entropy as functions
of a related i.i.d.\ source entropy, the alphabet size, the letter probabilities,
and their arrangement
in the probability space.  The bounds provided
a range of values the pattern entropy can take, and showed that in many cases
it \emph{must\/} decrease substantially from the original i.i.d.\ sequence entropy.
It was shown that low probability symbols contribute mostly as a single
super-symbol to the pattern entropy, where in particular, very low probability
symbols contribute negligibly over the contribution of this super-symbol.

\end{document}